\documentclass[eqsecnum,showpacs,aps]{revtex4}
\usepackage{epsfig}

\newcommand{\beq}{\begin{equation}}
\newcommand{\eeq}{\end{equation}}
\newcommand{\beqa}{\begin{eqnarray}}
\newcommand{\eeqa}{\end{eqnarray}}

\def\ket#1{|\,#1\,\rangle}
\def\bra#1{\langle\, #1\,|}

\def\proj#1#2{\ket{#1}\bra{#2}}

\def\ol#1{\overline{#1}}

\begin{document}
\title{Asymmetric two-output quantum processor in any dimension}

\author{Iulia Ghiu $^{1,2}$}
\email{iughiu@barutu.fizica.unibuc.ro}

\author{Gunnar Bj\"ork $^2$}
\email{gunnarb@imit.kth.se}

\affiliation{$^1$Department of Physics, Section of Quantum Mechanics
and Statistical Physics, University of Bucharest, P.O. Box MG-11,
R-077125, Bucharest-M\u{a}gurele, Romania}

\affiliation{$^2$School of Information and Communication Technology,
Royal Institute of Technology (KTH), Electrum 229, SE 164 40 Kista,
Sweden}

\date{\today}

\begin{abstract}
We propose two different implementations of an asymmetric
two-output probabilistic quantum processor, which can implement a
restricted set of one-qubit operations. One of them is constructed by
combining asymmetric telecloning with a quantum gate array. We
analyze the efficiency of this processor by evaluating the
fidelities between the desired operation and the one generated by
the processor and show that the two output states are the same
as the ones produced by the optimal universal asymmetric Pauli
cloning machine. The schemes require only local operations
and classical communication, they have the advantage of transmitting
the two output states directly to two spatially separated receivers
but they have a success probability of 1/2. We show further that we
can perform the same one-qubit operation with unity probability
at the cost of using nonlocal operations. We finally
generalize the two schemes for $D$-level systems and find that the
local ones are successful with a probability of $1/D$ and the
nonlocal generalized scheme is always successful.

 \end{abstract}

\maketitle

\section{Introduction}

Quantum computers are machines that employ quantum phenomena,
such as quantum interference and entanglement, to solve a particular
problem. The computers have to execute a program, which is built
with the help of a precise set of instructions, in order to give the
desired solution. A specification of this set of instructions is
called an algorithm. There are algorithms which can be performed
faster on a quantum computer than on a classical one
\cite{Nielsen1}: the Deutsch-Jozsa algorithm for solving the
oracle problem \cite{Deutsch1,Deutsch2,exp}, the Shor
algorithm for factoring large integers \cite{Shor}, and the
Grover algorithm for searching unsorted databases \cite{Grover}.

The most important component of the computer architecture is
the processor. One crucial property of the classical processor is
that we keep the same circuit regardless of the instructions that we
want to perform. One may then ask how to construct a universal
quantum processor, a fixed device that implements any desired
program on the information stored in quantum systems? This problem
was originally investigated by Nielsen and Chuang \cite{Nielsen2},
where they proposed a model of the quantum processor, which consists
of a quantum gate array $G$ acting on the data state $\ket{\psi}_d$
and on the program state $\ket{P}$. The dynamics of the quantum gate
array is 
\beq 
G[\ket{d}\ket{P_U}]=(U\ket{d})\ket{P'_U}, 
\eeq 
where $U$ is a particular unitary operator implemented by the processor.
Nielsen and Chuang found two important results \cite{Nielsen2}: (i)
the state $\ket{P'_U}$ is independent of the data register $\ket{d}$
and (ii) no deterministic universal quantum gate array exists.
Therefore they showed how to construct an one-output probabilistic
quantum processor, whose operating principle is that of
quantum teleportation \cite{Bennett}. The outcome of a Bell
measurement tells us when the desired operation succeeded.

In the last few years, much progress has been made on
generalizations and applications of the probabilistic quantum
processor. Huelga {\it et al.} found a generalization of the
method of teleporting a quantum gate from one location to another
\cite{Huelga1,Huelga2,Huelga3}. Two more proposals of probabilistic
quantum processors were considered by Preskill
\cite{Preskill} and Vidal {\it et al.} \cite{Vidal}. Vidal {\it et
al.} analyzed a probabilistic gate, which performs an arbitrary
rotation around the $z$ axis of a spin-1/2 particle with the help of
an $N$-qubit program state \cite{Vidal}. More complex probabilistic
programmable quantum processors have been proposed by Hillery {\it
et al.} \cite{Hillery1,Hillery3} by investigating the case
when an arbitrary linear operation $A$ is performed. They have built
the network for this processor by using a quantum information
distributor \cite{Braunstein,Rosko} for qubits and then for
$D$-level systems (quDit). Hillery {\it et al.} have analyzed
several classes of quantum processors, which execute more general
operations, namely completely positive maps, on quantum systems
\cite{Hillery4}. In addition, they have found two important results:
one can build a quantum processor to perform the phase-damping
channel and that this is not possible in the case of the
amplitude-damping channel \cite{Hillery4}. 

Further extensions have been developed. In Refs.
\cite{Hillery2,Hillery5} a quantum processor, which executes
SU(N) rotations was considered, and was found that the probability
of success for implementing the operation is increased if
conditionated loops are used. Recently Brazier {\it et al.} have
investigated the case when we have access to many copies of the
program state \cite{Brazier}. They have shown that the probability
of success cannot be increased and that it is the same as the one
obtained using two different schemes: VMC \cite{Vidal} and HZB
\cite{Hillery2}. Positive-operator-valued measures (POVM) are the
most general measurements allowed by quantum mechanics
\cite{Peres,Nielsen1}. Therefore it would be interesting to study
the possibility of realizations of POVMs on quantum processors. This
problem was investigated by Ziman and Bu\v{z}ek \cite{Ziman}, where
they showed how to encode a POVM into a program state.
Another important class of operations is the one of generators of
Markovian dynamics, which are relevant in the context of quantum
decoherence. Koniorczyk {\it et al.} have recently proposed a
scenario for the simulation of the infinitesimal generators of the
Markovian semigroup on quantum processors \cite{Koniorczyk}.
Approximate processors, i.e. processors which implement a set of
unitary operators with high precision, have been introduced by
Hillery {\it et al.} in Ref. \cite{Hillery6}. The accuracy of the
processor is given by the process fidelity, which was shown to be
maximum if one chooses the program state to be the eigenvector
corresponding to the largest eigenvalue of a certain operator. This
operator depends on some operators $A_{jk}$, which characterize the
quantum processor, and the desired unitary operator $U$ to be
implemented. We emphasize that all the processors
described above generate only one output state. A two-output
processor was proposed by Yu {\it et al.}  \cite{Yu} by combining
symmetric telecloning of qubits \cite{Murao} with a programmable
quantum gate array.

In this paper we present two different schemes for obtaining
a two-output quantum gate for $D$-level systems. Suppose the
following scenario: an observer Peter has to teleport the result of
a certain operation to two distant parties, Alice and Bob. We show
how this task can be accomplished using a shared entangled
state, local operations, and classical communication (LOCC) by
proposing two schemes. We restrict our study to a certain class of
unitary operations, which depends on a parameter $\theta \in [0,2\pi
)$. In the first scheme, described in Sec. II. A, we extend
Yu {\it et al.}'s scheme by considering asymmetric telecloning
of qubits. The program state consists of a
four-particle entangled state shared between Peter, Alice, Bob, and
another observer Charlie, who holds an ancillary system. (As
Charlie's particle is only used as a resource, and is discarded at
the end, Peter and Charlie may be the same party in those schemes
that do not need nonlocal operations between Alice's, Bob's and
Charlie's particles.) Peter measures the data qubit and his
particle $P$, which is included in the program register, in the
standard Bell basis, and then communicates the outcome. With a
probability of 1/2, Alice and Bob are able to recover the desired
mixed output states, whose fidelities are identical with the ones
obtained by the optimal universal asymmetric Pauli cloning machine.
We also demonstrate that if Alice, Bob, and Charlie may use
nonlocal operations, the protocol will always be successful with
preserved fidelities. In Sec. II. B, we propose the second local
protocol, which requires a four-particle entangled state as a
quantum channel being distributed between Peter, Alice, Bob, and
Charlie. The desired unitary operation is encoded in a generalized
Bell basis. The protocol is as follows: Peter performs a measurement
in the generalized Bell basis and then announces the outcome to
Alice, Bob, and Charlie. The probability of success of this second
scheme is 1/2 and the asymmetric outputs received by Alice and Bob
are identical with the ones obtained in the previous scheme. Given
the resources of the four-particle states, these local schemes
present the advantage that they can be implemented in the lab. 

In Sec. III we present the generalizations of these protocols for
quDits. The two local generalized schemes are successful with a
probability equal to $1/D$, while unit probability of success
again requires nonlocal operations. Finally, in Sec. IV we
summarize our conclusions.

%.............................................................
\section{Asymmetric quantum gate array for qubits}

In this section we present two schemes for performing the following
scenario: we start with an arbitrary qubit state 
\beq
\ket{\psi}_d=\alpha _0\ket{0}+\alpha _1\ket{1} , \label{qubit state}
\eeq 
where $|\alpha_0|^2+|\alpha_1|^2=1$.  
We want to obtain
two optimal universal asymmetric clones of a certain unitary
computational operator described by:
\beq \label{uteta}
U_\theta=\left( \begin{array}{cc}
e^{i\theta }&0\\
0&e^{-i\theta }
\end{array}\right),
\eeq 
with $\theta \in [0,2\pi )$, applied on the arbitrary input
data state $\ket{\psi}_d$.

\subsection{Two-output quantum processor for qubits}

We define a state required in the preparation of the program as \beq
\label{csi} \ket{\xi}_{PABC}=\frac{1}{\sqrt 2}\left(
\ket{0}_P\ket{\phi _0}_{ABC}+\ket{1}_P\ket{\phi _1}_{ABC}\right),
\eeq 
where 
\beqa
\ket{\phi _0}_{ABC}&=&\frac{1}{\sqrt{2(1-p+p^2)}}\left[ \ket{000}+p\ket{011}+(1-p)\ket{101}\right], \hspace{0.2cm}\mbox{and}\nonumber\\
\ket{\phi _1}_{ABC}&=&\frac{1}{\sqrt{2(1-p+p^2)}}\left[
\ket{111}+p\ket{100}+(1-p)\ket{010}\right], 
\eeqa 
with $0<p<1$. The
two states $\ket{\phi_0}$ and $\ket{\phi_1}$ are obtained by
applying an optimal universal asymmetric Pauli cloning machine on
the states $\ket{0}\ket{00}$ and $\ket{1}\ket{00}$ \cite{Ghiu}. The
data system $d$ and the first qubit $P$ of the state $\ket{\xi}$
belong to an observer Peter, while the qubits $A, B, C$ are held by
other distant parties, Alice, Bob, and Charlie, respectively.
This ``ownership'' of the qubits (or, later, quDits) will
hold for all our schemes. Note that, in the case of the ``local''
schemes, Peter and Charlie may be identically the same, as Charlie's
qubit is discarded at the end.

Peter wants to teleport the result of the desired unitary operator
$U_\theta $ to Alice and Bob. Peter encodes the information carried
by the unitary operator, in the program state $\ket{P_U}_{PABC}$ by
locally applying $U_\theta $ on the qubit $P$ (see Fig. 1):  
\beq
\ket{P_U}_{PABC}=U_\theta\otimes
I_{ABC}\ket{\xi}_{PABC}=\frac{1}{\sqrt 2}\left( e^{i\theta
}\ket{0}_P\ket{\phi_0}_{ABC}+e^{-i\theta
}\ket{1}_P\ket{\phi_1}_{ABC}\right). 
\eeq

Let us write now the input state with the help of the standard Bell basis
$\ket{\Phi^{\pm}}=\frac{1}{\sqrt 2}(\ket{00}\pm \ket{11})$ and
$\ket{\Psi^{\pm}}=\frac{1}{\sqrt 2}(\ket{01}\pm \ket{10})$:
\beqa
\ket{\psi}_d\ket{P_U}_{PABC}&=&\frac{1}{2}[\ket{\Phi^+}_{dP}(\alpha_0e^{i\theta }\ket{\phi_0}+\alpha_1e^{-i\theta }\ket{\phi_1})\nonumber\\
&&+\ket{\Phi^-}_{dP}(\alpha_0e^{i\theta }\ket{\phi_0}-\alpha_1e^{-i\theta }\ket{\phi_1})\nonumber\\
&&+\ket{\Psi^+}_{dP}(\alpha_1e^{i\theta }\ket{\phi_0}+\alpha_0e^{-i\theta }\ket{\phi_1})\nonumber\\
&&+\ket{\Psi^-}_{dP}(\alpha_1e^{i\theta
}\ket{\phi_0}-\alpha_0e^{-i\theta }\ket{\phi_1})]. 
\eeqa
Peter performs a measurement in the Bell basis on the data
qubit and the first qubit $P$ in the program state as it is shown in
Fig. 1. Then he communicates the outcome of the measurement to
Alice, Bob, and Charlie. With a probability equal to $1/4$ the
outcome is $\ket{\Phi^+}_{dP}$ and therefore the output is projected
to 
\beq \label{steta} 
\ket{\eta }_{ABC}=\alpha_0e^{i\theta
}\ket{\phi_0}+\alpha_1e^{-i\theta }\ket{\phi_1}. 
\eeq

\begin{figure}
\includegraphics[height=5cm,width=12cm]{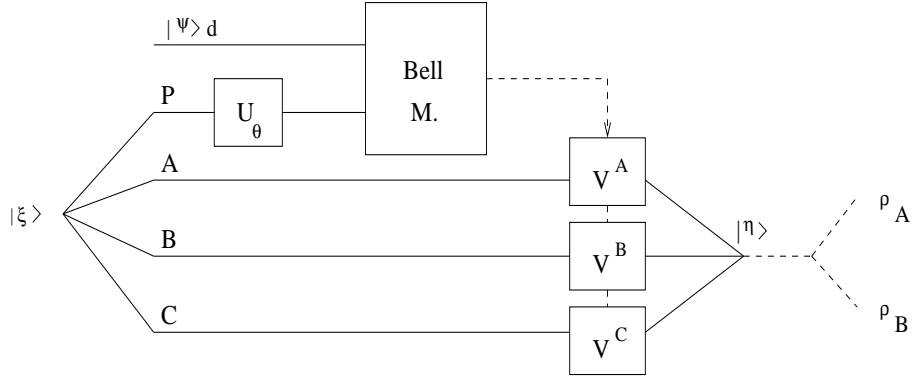}
\vspace{0.5cm} \caption{The scheme for the quantum processor for
qubits. The input states consist of: an arbitrary data
register $\ket{\psi}_d$ and the program register
$\ket{P_U}_{PABC}=U_\theta \otimes I_{ABC}\ket{\xi }_{PABC}$. Alice
and Bob will obtain two mixed states $\rho_A, \rho_B$ that are
implemented by the quantum processor. }  
\end{figure}
Accordingly, after tracing over Charlie's qubit, the two
final states of Alice and Bob, rspectively, are:
 \beqa
\label{final}
\rho_A&=&\mbox{Tr}_{B,C}\proj{\eta}{\eta}=\frac{1}{2(1-p+p^2)}\{[2p|\alpha_0|^2+(1-p)^2]\proj{0}{0}+[2p|\alpha_1|^2+(1-p)^2]\proj{1}{1}]\nonumber\\
&&+2p\alpha_0\alpha_1^*e^{2i\theta }\proj{0}{1}+2p\alpha_0^*\alpha_1e^{-2i\theta }\proj{1}{0}\};\nonumber\\
\rho_B&=&\mbox{Tr}_{A,C}\proj{\eta}{\eta}=\frac{1}{2(1-p+p^2)}\{[2(1-p)|\alpha_0|^2+p^2]\proj{0}{0}+[2(1-p)|\alpha_1|^2+p^2]\proj{1}{1}]\nonumber\\
&&+2(1-p)\alpha_0\alpha_1^*e^{2i\theta
}\proj{0}{1}+2(1-p)\alpha_0^*\alpha_1e^{-2i\theta }\proj{1}{0}\}.
\eeqa
The efficiency of the quantum processor is evaluated with the
help of the fidelities of the output states with respect to the
exact data register outputs $U_\theta \ket{\psi}_d$: 
\beqa
\label{fid-proc}
F_A&=&_d\bra{\psi}U_\theta ^\dagger \rho_AU_\theta \ket{\psi}_d=\frac{1+p^2}{2(1-p+p^2)}, \hspace{0.2cm}\mbox{and}\nonumber\\
F_B&=&_d\bra{\psi}U_\theta ^\dagger \rho_BU_\theta \ket{\psi}_d=\frac{2-2p+p^2}{2(1-p+p^2)}.
\eeqa

If the output of the measurement is $\ket{\Phi^-}_{dP}$, then the
final state is 
\beq 
\ket{\eta '}_{ABC}=\alpha_0e^{i\theta
}\ket{\phi_0}-\alpha_1e^{-i\theta }\ket{\phi_1}. 
\eeq 
Alice, Bob, and Charlie can transform the state $\ket{\eta '}$ to the
state $\ket{\eta}$ of Eq. (\ref{steta}) by applying the local
unitary operator $V=\sigma_z^A\otimes \sigma_z^B\otimes \sigma_z^C$.
Therefore, in this case, Alice and Bob obtain the same final
states $\rho_A$ and $\rho_B$ as above. For the other two outcomes,
when Peter obtains $\ket{\Psi^+}_{dP}$ and
$\ket{\Psi^-}_{dP}$, the result cannot be transformed to the state
of Eq. (\ref{steta}) by local operations. Hence there is no
chance to obtain the desired mixed outputs $\rho_A$ and $\rho_B$
if we don't allow for nonlocal transformations. The processor
hence succeeds with the probability $1/2$ and generates two
asymmetric output mixed states. The fidelities of the final states
given by Eq. (\ref{fid-proc}) are identical with the ones of the
clones emerging from the optimal universal asymmetric cloning
machine given in Refs. \cite{Ghiu,Cerf1,Cerf2}. In the particular
case when $p=1/2$, we recover the result of Yu {\it et al.}
\cite{Yu}.

However, if we allow nonlocal operations between Alice, Bob,
and Charlie, which either entails their respective particles to
interact directly, or via a shared auxiliary entangled state, it is
also possible to convert the states $\ket{\Psi^+}_{dP}$ and
$\ket{\Psi^-}_{dP}$ to $\ket{\eta}$ and thereby always succeed with
the protocol. This is generally true for all our proposed schemes.

%.........................................................
\subsection{Two-output local quantum gate for qubits}

We use the same channel $\ket{\xi }_{PABC}$ of Eq. (\ref{csi})
initially shared by Peter, Alice, Bob, and Charlie as above,
but this time Peter performs a measurement in a different Bell
basis, which depends on $\theta $ (see Fig. 2): 
\beqa
\ket{\tilde \Phi ^{\pm }}&=&\frac{1}{\sqrt 2}\left( e^{-i\theta }\ket{00}\pm e^{i\theta }\ket{11}\right),\nonumber\\
\ket{\tilde \Psi ^{\pm }}&=&\frac{1}{\sqrt 2}\left( e^{-i\theta }\ket{01}\pm e^{i\theta }\ket{10}\right) .
\eeqa
The input state can be written as follows
\beqa
\ket{\psi}_d\ket{\xi }_{PABC}&=&\frac{1}{2}[\ket{\tilde \Phi^+}_{dP}(\alpha_0e^{i\theta }\ket{\phi_0}+\alpha_1e^{-i\theta }\ket{\phi_1})\nonumber\\
&&+\ket{\tilde \Phi^-}_{dP}(\alpha_0e^{i\theta }\ket{\phi_0}-\alpha_1e^{-i\theta }\ket{\phi_1})\nonumber\\
&&+\ket{\tilde \Psi^+}_{dP}(\alpha_1e^{i\theta }\ket{\phi_0}+\alpha_0e^{-i\theta }\ket{\phi_1})\nonumber\\
&&+\ket{\tilde \Psi^-}_{dP}(\alpha_1e^{i\theta
}\ket{\phi_0}-\alpha_0e^{-i\theta }\ket{\phi_1})]. 
\eeqa 
Peter communicates the outcome to Alice, Bob, and Charlie. If the
result of the measurement of particles $d$ and $P$ is $\ket{\tilde
\Phi^+}_{dP}$, then we get the same state $\ket{\eta }$ of Eq.
(\ref{steta}): 
\beq 
\ket{\eta }_{ABC}=\alpha_0e^{i\theta
}\ket{\phi_0}+\alpha_1e^{-i\theta }\ket{\phi_1}. 
\eeq 
Alice and Bob
easily obtain the two asymmetric outputs $\rho _A$ and $\rho_B$
given by Eq. (\ref{final}). If the outcome is $\ket{\tilde
\Phi^-}_{dP}$, then Alice, Bob, and Charlie apply the local
operator $V=\sigma_z^A\otimes \sigma_z^B\otimes \sigma_z^C$ in order
to recover the state $\ket{\eta}$. The other two outcomes
$\ket{\tilde \Psi^+}_{dP}$ and $\ket{\tilde \Psi^-}_{dP}$ will lead
to a different result and the procedure fails unless nonlocal
transformations are used. Hence, the total success probability is
$p=1/2$. This protocol is constructed using only LOCC and therefore
it is suitable for distant computation at two locations.

\begin{figure}
\includegraphics[height=5cm,width=12cm]{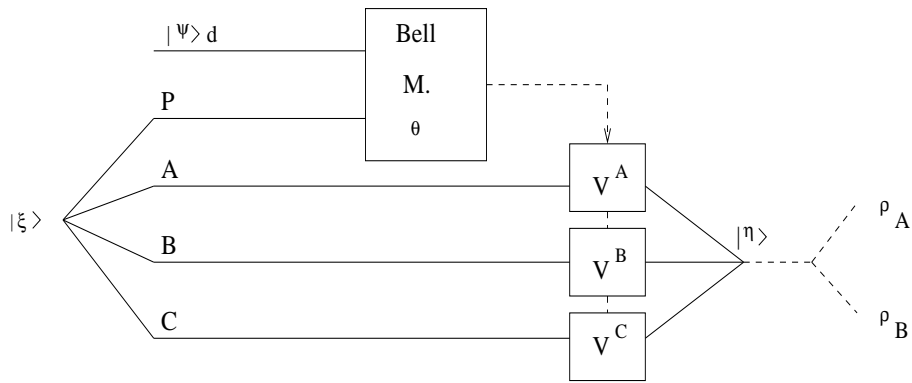}
\vspace{0.5cm} \caption{The scheme for the local quantum gate-array
for qubits. The information of the operation to be performed
is encoded in a generalized Bell basis. Peter performs a measurement
in this generalized Bell basis and then communicates the outcome to
Alice, Bob, and Charlie. By LOCC, Alice and Bob are able to get the
mixed states $\rho_A, \rho_B$. }
\end{figure}

%.........................................................

\section{Asymmetric quantum gate array for quDits}

\subsection{Two-output quantum processor for quDits}

One-output probabilistic programable quantum processors have
recently been analyzed by Hillery {\it et al.}
\cite{Hillery1,Hillery2} in the case when the data are encoded on a
$D$-level systems (quDits). More precisely, they have investigated
the possibility to construct a processor which performs an arbitrary
linear operation $A$ \cite{Hillery1}: 
\beq
A=\sum_{m,n=0}^{D-1}q_{mn}U^{(mn)}, 
\eeq where the operators
$U^{(mn)}$ form a basis in the Hilbert-Schmidt space 
\beq
U^{(mn)}:=\sum_{s=0}^{D-1}\mbox{exp}\left( -\frac{2\pi
ism}{D}\right) \proj{s-n}{s}. 
\eeq

We now propose a generalization of the quantum processor presented
in the Sec. II. A for quDits. An arbitrary data quDit-state is
described by 
\beq \label{cunit}
\ket{\psi}_d=\sum_{k=0}^{D-1}\alpha_k\ket{k}, 
\eeq where
$\sum_{k=0}^{D-1}|\alpha_k|^2=1$. We want to analyze the action of
the restricted class of unitary operators given by 
\beq
U_\theta=\mbox{cos}\theta \;I+i\; \mbox{sin}\theta \; U^{\left(
\frac{D}{2},0\right) }= \sum_{s=0}^{D-1}\mbox{exp}[(-1)^si\theta
]\proj{s}{s}= \left( \begin{array}{ccccc}
e^{i\theta }&0&\ldots &0&0\\
0&e^{-i\theta }&\ldots &0&0\\
\vdots &\vdots &\ddots &\vdots &\vdots \\
0&0&\ldots &e^{i\theta }&0\\
0&0&\ldots &0&e^{-i\theta }
\end{array}\right),
\eeq
where $D$ is assumed to be even. We define a family of states which depends on a parameter $p$, $0<p<1$:
\beqa \label{general}
\ket{\phi_j}
&=&\frac{1}{\sqrt{1+(D-1)(2p^2-2p+1)}}[ \ket{j}\ket{j}\ket{j}+p\sum_{r=1}^{D-1}\ket{j}\ket{\ol{j+r}}\ket{\ol{j+r}}\nonumber\\
&&+(1-p)\sum_{r=1}^{D-1}\ket{\ol{j+r}}\ket{j}\ket{\ol{j+r}}], \eeqa
where $j=0,..., D-1$, and $\ol{j+r}=j+r$ modulo $D$. These states
were found by one of us in Ref. \cite{Ghiu} by considering the
action of an optimal universal asymmetric Heisenberg cloning machine
on the state $\ket{j}\ket{00}$. In addition, we introduce the state
$\ket{\xi}_{PABC}$ as follows: 
\beq \label{xiN}
\ket{\xi}_{PABC}=\frac{1}{\sqrt
D}\sum_{j=0}^{D-1}\ket{j}_P\ket{\phi_j}_{ABC}. 
\eeq 
Peter encodes the operation $U_\theta $ in the state of a program
register $\ket{P_U}_{PABC}$ as follows 
\beq
\ket{P_U}_{PABC}=U_\theta \otimes I_{ABC}\ket{\xi}=\frac{1}{\sqrt
D}\sum_{j=0}^{D-1}\mbox{exp}[(-1)^ji\theta
]\ket{j}_P\ket{\phi_j}_{ABC}.
 \eeq
We denote by $\ket{\Phi_{m,n}}$
the standard Bell basis for quDits:
\beq
\ket{\Phi_{m,n}}=\frac{1}{\sqrt
D}\sum_{k=0}^{D-1}\mbox{exp}\left( \frac{2\pi ikn}{D}\right)
\ket{k}\ket{\ol{k+m}}.
\eeq
The input state can be written by using
the standard Bell basis as
\beq
\ket{\psi
}_d\ket{P_U}_{PABC}=\frac{1}{D}\sum_{k,m,n=0}^{D-1}\alpha_k\mbox{exp}[(-1)^{\ol
{k+m}}i\theta ]\mbox{exp}\left( -\frac{2\pi ikn}{D}\right)
\ket{\Phi_{m,n}}_{dP}\ket{\phi_{\ol{k+m}}}_{ABC}.
 \eeq
The desired
output state should be
\beq
\ket{\Lambda}=U_\theta
\ket{\psi}_d=\sum_{j=0}^{D-1}\alpha _j\mbox{exp}\left[ (-1)^ji\theta
\right]\ket{j}.
\eeq
A measurement in the standard Bell basis
onto the data state and the first qubit $P$ of the program state is
performed by Peter. Then he announces the outcome of the measurement to Alice, Bob, and Charlie. With a probability $1/D^2$ he gets the outcome
$\ket{\Phi_{0,n}}$ (where $n$ is a fixed number) and at the same
time the state of the quDits $A,B$ and $C$ becomes
\beq
\label{etaprim} \ket{\eta '}_{ABC}=\sum_{k=0}^{D-1}\alpha
_k\mbox{exp}\left[ (-1)^ki\theta \right]\mbox{exp}\left( -\frac{2\pi
ikn}{D}\right) \ket{\phi _k}_{ABC}.
\eeq

Let us now define a local operator $V_n:=V^A_n\otimes V^B_n\otimes V^C_n$, where
\beqa
V^X_n&:=&\sum_{j=0}^{D-1}\mbox{exp}\left( \frac{2\pi ijn}{D}\right)\proj{j}{j}, \hspace{1cm}X=A,B;\nonumber\\
V^C_n&:=&\sum_{j=0}^{D-1}\mbox{exp}\left( -\frac{2\pi
ijn}{D}\right)\proj{j}{j}. 
\eeqa 
Depending on Peter's outcome,
Alice, Bob, and Charlie apply subsequently the local operator $V_n$
on the output state (\ref{etaprim}) and obtain 
\beq \label{eta-gen}
V_n\ket{\eta '}_{ABC}=\ket{\eta}_{ABC}=\sum_{k=0}^{D-1}\alpha
_k\mbox{exp}\left[ (-1)^ki\theta \right] \ket{\phi _k}_{ABC}. 
\eeq
Since there are $D$ equiprobable measurement outcomes
$\ket{\Phi_{0,n}}$ for $n=0,..., D-1$, the total success probability
is $1/D$. The two mixed output states of the quantum processor are,
in all the $D$ cases, 
\beqa \label{roN}
\rho_A&=&\mbox{Tr}_{B,C}\proj{\eta}{\eta}=\frac{1}{1+(D-1)(2p^2-2p+1)}\bigg( \sum_{j=0}^{D-1}\left\{ \left[ 2p+(D-2)p^2\right] |\alpha_j|^2+(1-p)^2\right\}\proj{j}{j}\nonumber\\
&&+\left[ 2p+(D-2)p^2\right] \sum^{D-1}_{\stackrel{\textstyle{j,k=0}}{j\not= k}}\alpha _j\alpha _k^*\mbox{exp}\left\{ \left[ (-1)^j+(-1)^{k+1}\right] i\theta \right\} \proj{j}{k} \bigg);\nonumber\\
\rho_B&=&\mbox{Tr}_{A,C}\proj{\eta}{\eta}=\frac{1}{1+(D-1)(2p^2-2p+1)}\bigg( \sum_{j=0}^{D-1}\left\{ \left[ D-2(D-1)p+(D-2)p^2\right] |\alpha_j|^2+p^2\right\}\proj{j}{j}\nonumber\\
&&+\left[ D-2(D-1)p+(D-2)p^2\right] \sum^{D-1}_{\stackrel{\textstyle{j,k=0}}{j\not= k}}\alpha _j\alpha _k^*\mbox{exp}\left\{ \left[ (-1)^j+(-1)^{k+1}\right] i\theta \right\} \proj{j}{k} \bigg).
\eeqa
The fidelities of the output states are given by:
\beqa \label{fids}
F_A&=&\bra{\Lambda }\rho_A\ket{\Lambda }=\frac{1+(D-1)p^2}{1+(D-1)(2p^2-2p+1)},\hspace{0.2cm}\mbox{and}\nonumber\\
F_B&=&\bra{\Lambda }\rho_B\ket{\Lambda
}=\frac{1+(D-1)(1-p)^2}{1+(D-1)(2p^2-2p+1)}. 
\eeqa 
They are
identical with the ones generated by the optimal universal
asymmetric Heisenberg cloning machine given in Ref. \cite{Ghiu}.
In the case when the result of Peter's measurement is not
$\ket{\Phi_{0,n}}$, Alice, Bob, and Charlie are not able to recover
the state $\ket{\eta}_{ABC}$ given by Eq. (\ref{eta-gen}) only by
LOCC, and then the computation fails. However, both in this
and the following protocol, the success probability can be boosted
to unity if the three parties may use nonlocal operations on their
particles.

%..........................................................

\subsection{Two-output local quantum gate for quDits}

Our input data state is an arbitrary quDit given by Eq.
(\ref{cunit}) and belongs to Peter. We define a different
Bell basis for quDits depending on $\theta $: 
\beq 
\ket{\tilde
\Phi_{m,n}}=\frac{1}{\sqrt D}\sum_{k=0}^{D-1}\mbox{exp}\left[
(-1)^{k+1}i\theta \right] \mbox{exp}\left( \frac{2\pi ikn}{D}\right)
\ket{k}\ket{\ol{k+m}}. 
\eeq 
The input state can be written with the
help of this new basis as: 
\beq 
\ket{\psi }_d\ket{\xi
}_{PABC}=\frac{1}{D}\sum_{k,m,n=0}^{D-1}\alpha_k\mbox{exp}[(-1)^{\ol
{k+m}}i\theta ]\mbox{exp}\left( -\frac{2\pi ikn}{D}\right)
\ket{\tilde \Phi_{m,n}}_{dP}\ket{\phi_{\ol{k+m}}}_{ABC}, 
\eeq 
where
the state $\ket{\xi }_{PABC}$ was introduced in Sec. III. A by Eq.
(\ref{xiN}) while $\ket{\phi_j}$ are given by Eq. (\ref{general}).
Peter performs a measurement in the new Bell basis on
particles $d$ and $P$. The interesting measurement outcome
is $\ket{\tilde \Phi _{0,n}}$ obtained with the probability equal to
$1/D^2$. Accordingly, the projected state of particles $A$, $B$, and
$C$ is $\ket{\eta '}$ given by Eq. (\ref{etaprim}). Alice,
Bob, and Charlie have to apply the same local unitary operator
$V_n$ described in the previous subsection in order that Alice
and Bob to obtain the two asymmetric outputs $\rho _A$ and $\rho_B$
of Eq. (\ref{roN}). The total success probability of this scheme is
$1/D$, because there are $D$ outcome $\ket{\tilde
\Phi_{0,n}}$.

%..........................................................

\section{Conclusions}

In this paper we have demonstrated the possibility of
building a two-output programmable processor of qubits, which
generates two asymmetric states with a probability of success of
$1/2$. It is characterized by the same fidelities as the ones
generated by the optimal universal asymmetric Pauli cloning machine.
We have analyzed a second protocol which performs the same task,
being performed with the help of a generalized Bell basis. We
implement these schemes by using only local operations and classical
communication, and therefore they are suitable for distributed
computation to two spatially separated receivers.

In addition, we have presented the generalizations of these
two protocols to $D$-level systems. The generalizations present the
advantage of sending the outcomes to different locations and they
are achieved with a probability of 1/2.

Finally, we have shown how to increase the success
probability to unity by considering nonlocal operations. In this
case the two output states are obtained only after the particles A,
B, and C have interacted, either directly or through an auxiliary
entangled state.

\section*{Acknowledgments}
This work was sponsored by the Swedish Foundation for Strategic
Research (SSF), the Swedish Research Council (VR), the Romanian
CNCSIS through the grant TD 2915-232/2006 for the University of Bucharest, and CEEX for the University of Bucharest.

\end{document}